\documentclass[twocolumn,showpacs,preprintnumbers,amsmath,amssymb,pre]{revtex4}
\usepackage{graphicx}
\usepackage{dcolumn}
\usepackage{bm}

\begin{document}

\title{Critical behavior of self-assembled rigid rods on triangular and honeycomb lattices}

\author{L. G. L\'opez}
\author{D. H. Linares}
\author{A. J. Ramirez-Pastor}
\email{antorami@unsl.edu.ar} \affiliation{Departamento de
F\'{\i}sica, Instituto de F\'{\i}sica Aplicada, Universidad
Nacional de San Luis, CONICET, 5700 San Luis, Argentina}

\date{\today}

\begin{abstract}

Using Monte Carlo simulations and finite-size scaling analysis,
the critical behavior of self-assembled rigid rods on triangular
and honeycomb lattices at intermediate density has been studied.
The system is composed of monomers with two attractive (sticky)
poles that, by decreasing temperature or increasing density,
polymerize reversibly into chains with three allowed directions
and, at the same time, undergo a continuous isotropic-nematic IN
transition. The determination of the critical exponents, along
with the behavior of Binder cumulants, indicate that the IN
transition belongs to the $q=1$ Potts universality class.

\end{abstract}


\maketitle

\section{Introduction}

Self-assembly has been considered for over $50$ years to be
central to understanding structure formation in living systems
\cite{Workum}. As a consequence, an increasing interest has been
devoted to enhance our understanding of the theoretical basis of
the fundamental mechanisms governing self-assembly and the
observables required to characterize the interactions driving
thermodynamic self-assembly transitions
\cite{Teixeira,Lu,Zhang,Scior,Ouyang,Tavares0,Tavares,PRE4}.
Despite these studies, the knowledge of how this process works is
still incomplete and many of the basic principles characterizing
this type of organization are unclear.

It is obvious that a complete analysis of the self-assembly
phenomenon is a quite difficult subject because of the complexity
of the involved microscopic mechanisms. For this reason, the
understanding of simple models with increasing complexity might be
a help and a guide to establish a general framework for the study
of this kind of systems, and to stimulate the development of more
sophisticated models which can be able to reproduce concrete
experimental situations.

In this line of work, two previous articles \cite{Tavares,PRE4},
referred to as papers I and II, respectively, were devoted to the
study of a system of self-assembled rigid rods adsorbed on a
two-dimensional lattice. In paper I, Tavares et al. studied a
system composed of monomers with two attractive (sticky) poles
that polymerize reversibly into polydisperse chains and, at the
same time, undergo an isotropic-nematic (IN) continuous phase
transition. So, the interplay between the self-assembly process
and the nematic ordering is a distinctive characteristic of these
systems. Using an approach in the spirit of the Zwanzig model
\cite{Zwanzig}, the authors found that nematic ordering enhances
bonding. In addition, the average rod length was described
quantitatively in both phases, while the location of the ordering
transition, which was found to be continuous, was predicted
semiquantitatively by the theory. With respect to the
characteristics of the phase transition, Tavares et al. assumed as
working hypothesis that the nature of the IN transition remains
unchanged with respect to the case of monodisperse rigid rods on
square lattices, where the transition is in the 2D Ising
universality class \cite{GHOSH,EPL1,JCP7}.

Paper II was a step further, analyzing the universality class of
the IN phase transition at intermediate density. For this purpose,
an extensive work of Monte Carlo (MC) simulations and finite-size
scaling (FSS) analysis was carried out. The obtained results
showed that the self-assembly process affects the nature of the
transition. Thus, the accurate determination of the critical
exponents indicated that, for a square lattice, the universality
class of the IN transition changes from 2D Ising-type for
monodisperse rods without self-assembly \cite{EPL1} to $q=1$
Potts-type for self-assembled rods.

In this context, the objectives of the present paper are (1) to
extend the previous work to triangular and honeycomb lattices
using the same techniques developed in paper II; and (2) to study
the effect of the lattice structure on the critical behavior of
self-assembled rigid rods. For this purpose, MC simulations
\cite{BINDER} supplemented by analysis using FSS theory
\cite{PRIVMAN} have been carried out to study the critical
behavior in a system of self-assembled rigid rods deposited on
triangular and honeycomb lattices. As in paper II, the
calculations were developed at constant temperature and different
densities, being $(\theta/\theta_c-1)$ the normalized scaling
variable, where $\theta$ and $\theta_c$ represent density and
critical density, respectively. The accurate determination of the
critical exponents, along with the behavior of Binder cumulants,
confirmed that the IN transition of self-assembled rigid rods on
triangular and honeycomb lattices at intermediate density belongs
to the $q=1$ Potts universality class. The outline of the paper is
as follows. In Sec. II we describe the lattice-gas model and the
simulation scheme. In Sec. III we present the MC results and the
general conclusions.

\begin{figure*}[t]
\includegraphics[width=6cm,clip=true]{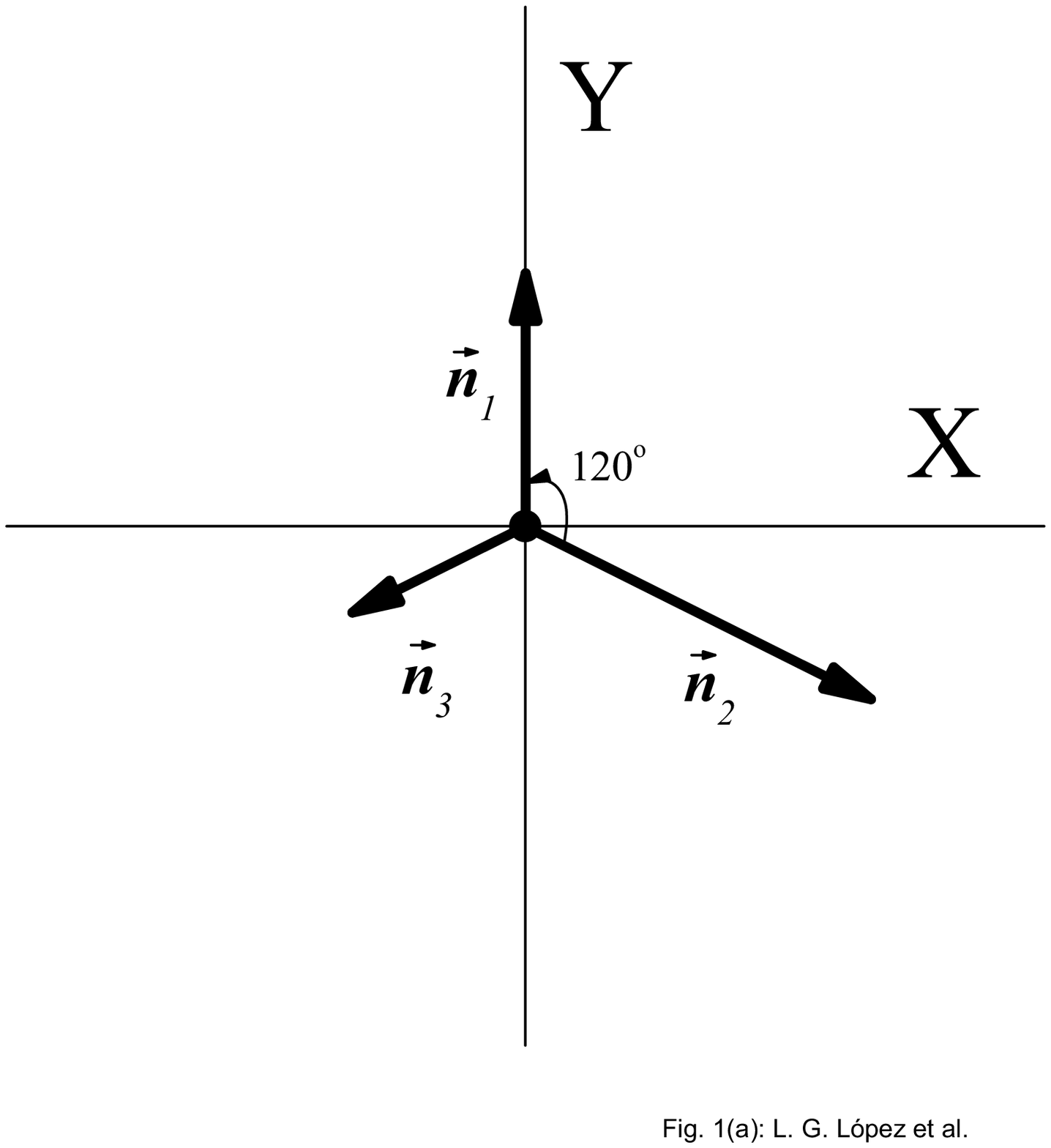}(a)
\includegraphics[width=6cm,clip=true]{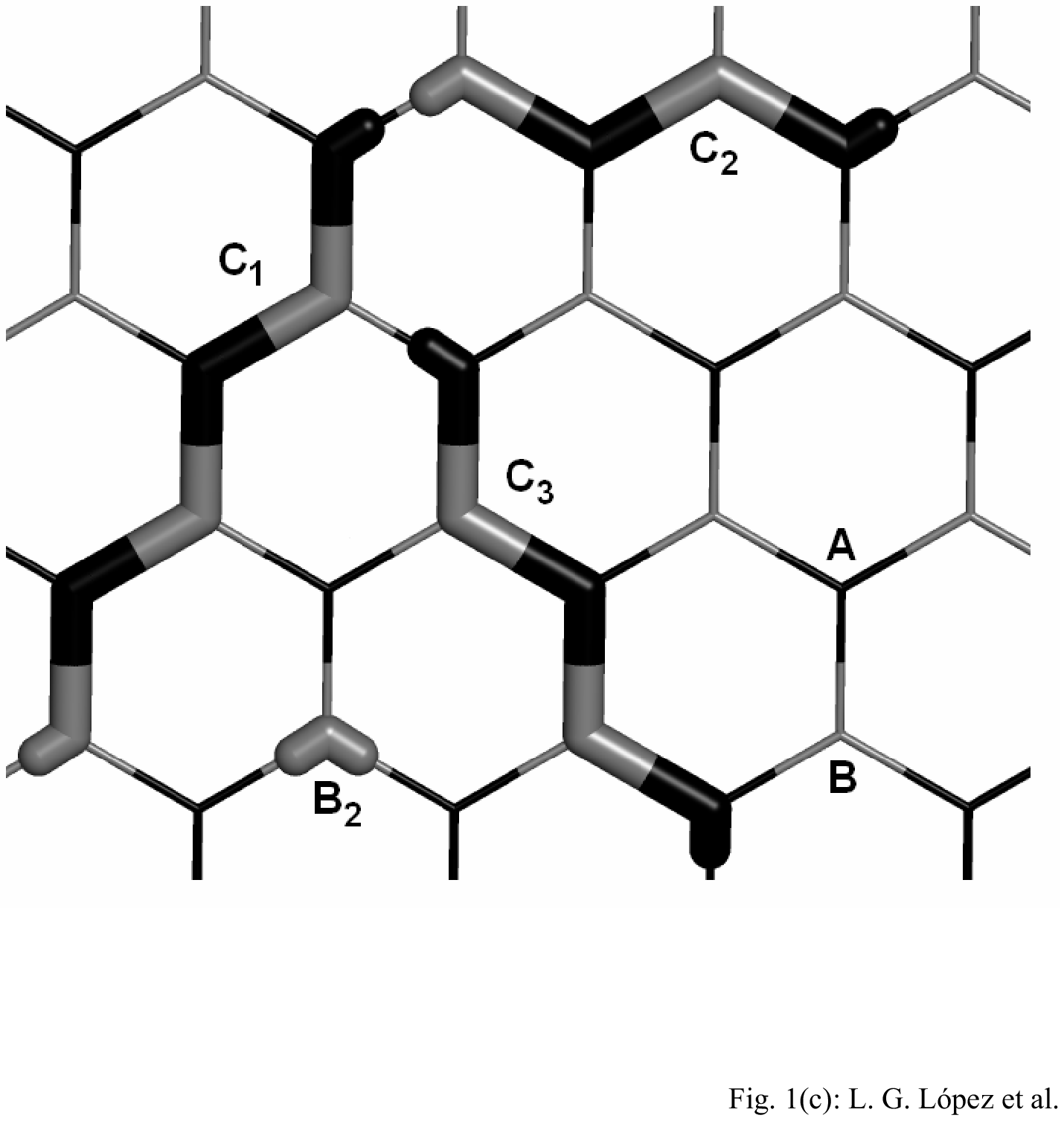}(c)
\includegraphics[width=6cm,clip=true]{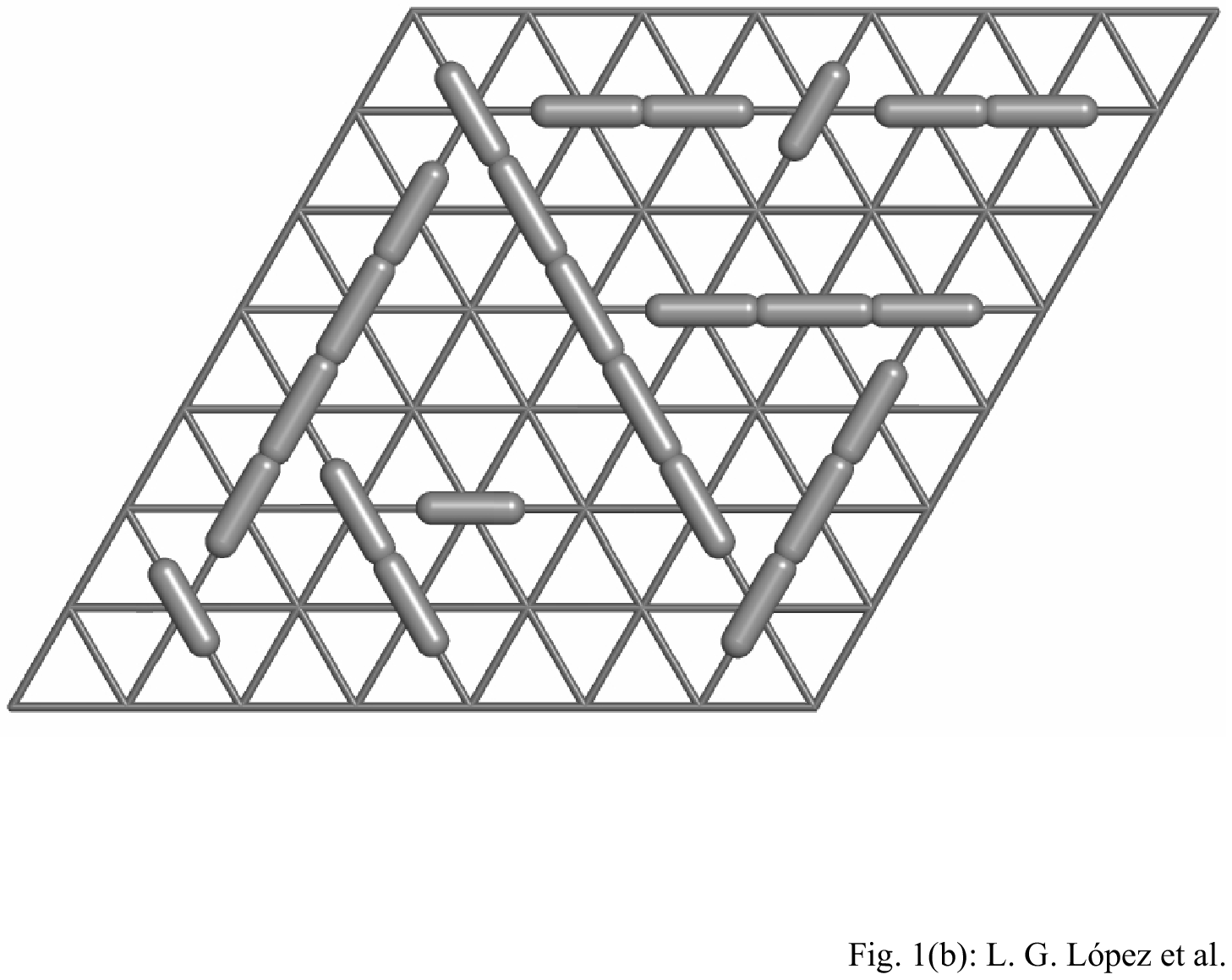}(b)
\includegraphics[width=6cm,clip=true]{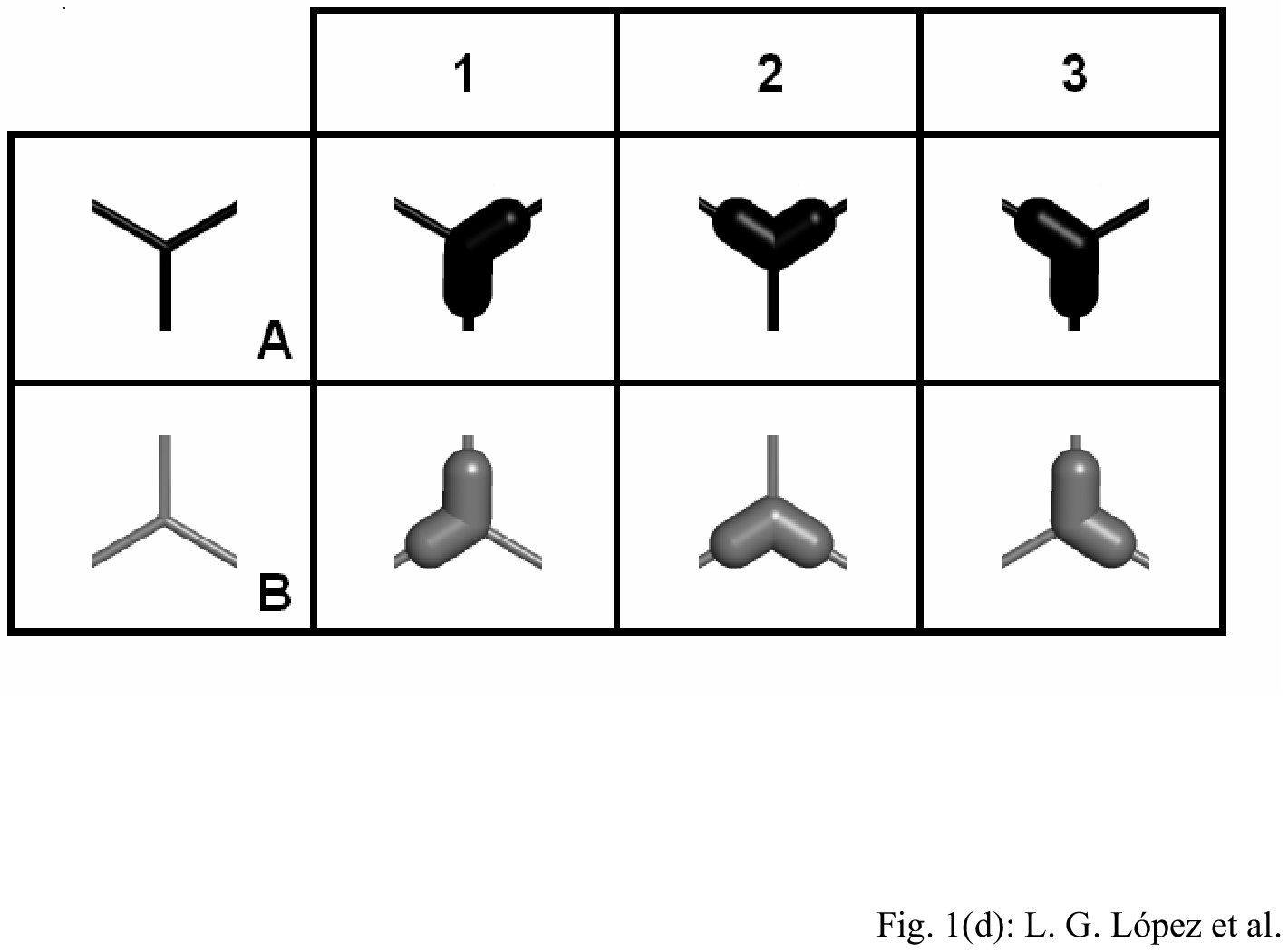}(d)
\caption{(a) Schematic representation of the set of vectors
$\{\vec{n}_1, \vec{n}_2, \vec{n}_3\}$ for triangular and honeycomb
lattices. (b) Self-assembled rigid rods on triangular lattices.
(c) Self-assembled rigid rods on honeycomb lattices. See
discussion in the text. (d) Available configurations for a monomer
on a honeycomb lattice. } \label{figure1}
\end{figure*}

\section{Lattice-gas model and Monte Carlo simulation scheme}

As in papers I and II, we consider a system of self-assembled rods
with a discrete number of orientations in two dimensions. In this
case, the surface is represented as an array of $M = L \times L$
adsorption sites in a triangular or in a honeycomb lattice
arrangement and periodic boundary conditions. $N$ particles are
adsorbed on the substrate with three possible orientations along
the principal axis of the lattice. These particles interact with
nearest-neighbors (NN) through anisotropic attractive
interactions. Thus, a cluster or uninterrupted sequence of bonded
particles is a self-assembled linear rod.

Under these considerations, the adsorbed phase is characterized by
the Hamiltonian
\begin{equation}
H = -w \sum_{\langle i,j \rangle} |\vec{r}_{ij} \cdot
\vec{\sigma}_i||\vec{r}_{ji} \cdot \vec{\sigma}_j| \ {\rm div} \
1, \label{h}
\end{equation}
where $\langle i,j \rangle$ indicates a sum over NN sites; ${\rm
div}$ means integer division; $w$ represents the NN lateral
interaction between two neighboring $i$ and $j$, which are aligned
with each other and with the intermolecular vector $\vec{r}_{ij}$;
and $\vec{\sigma}_i$ is the occupation vector with
$\vec{\sigma}_i=0$ if the site $i$ is empty, $\vec{\sigma}_i=
\hat{x_1}$ if the site $i$ is occupied by a particle with
orientation along the $x_1$-axis, $\vec{\sigma}_i= \hat{x_2}$ if
the site $i$ is occupied by a particle with orientation along the
$x_2$-axis, and $\vec{\sigma}_i= \hat{x_3}$ if the site $i$ is
occupied by a particle with orientation along the $x_3$-axis.

At fixed temperature, the average rod length increases as the
density increases and the polydisperse rods will undergo an
nematic ordering transition \cite{Tavares,PRE4}. In order to
follow the formation of the nematic phase, we use the order
parameter defined in previous work \cite{Tavares,PRE4}, which in
this case can be written as
\begin{equation}
\delta =  \frac{\left | \vec{n}_1 + \vec{n}_2 + \vec{n}_3 \right
|}{\left | \vec{n}_1 \right | + \left | \vec{n}_2 \right |+ \left
| \vec{n}_3 \right |}
 \label{fi}
\end{equation}
where each vector $\vec{n}_m$ is associated to one of the $3$
possible orientations (or directions) for a chain on the lattice.
In addition, $(1)$ the $\vec{n}_i$'s lie in the same plane (or are
co-planar) and point radially outward from a given point $P$ which
is defined as coordinate origin; $(2)$ the angle between two
consecutive vectors, $\vec{n}_i$ and $\vec{n}_{i+1}$, is equal to
$2\pi/3$; and $(3)$ the magnitude of $\vec{n}_i$ is equal to the
number of $k$-mers aligned along the $i$-direction [see Fig.
1(a)]. Note that the $\vec{n}_i$'s have the same directions as the
$q$ vectors in Ref.~[\onlinecite{WU}]. These directions are not
coincident with the allowed directions for the chains on the real
lattice.

The concept of linear rod is trivial for triangular lattices [see
Fig. 1(b)]. However, in a honeycomb lattice, the geometry does not
allow for the existence of a linear array of monomers. In this
case, we call ``linear rod" to a chain of adjacent monomers that
can be assembled in only three types of sequences, defining three
directions in similar way to the triangular lattice [see Fig.
1(c)]. Two ``types" of sites can be recognized in the honeycomb
lattice: $A$ (grey) and $B$ (black). Monomers (colored according
to the adsorption site) can be placed on these sites in three
different orientations: $A_1$, $A_2$, and $A_3$ (for the
$A$-sites); and $B_1$, $B_2$, and $B_3$ (for the $B$-sites) as
shown in Fig. 1(d). So, the self-assembled chains are formed by
alternating $A$ and $B$ monomers. In the particular case shown in
Fig. 1(c), $C_1$, $C_2$, and $C_3$ are built by the following
sequences: $B_1A_1B_1A_1B_1A_1$ ($C_1$), $B_2A_2B_2A_2$ ($C_2$),
and $A_3B_3A_3B_3A_3$ ($C_3$). The figure also shows an
unassociated $B$-monomer ($B_2$).

The problem has been studied by canonical Monte Carlo simulations
using an vacancy-particle-exchange Kawasaki dynamics
\cite{KAWASAKI} and Metropolis acceptance probability
\cite{METROPOLIS}. Typically, the equilibrium state can be well
reproduced after discarding the first $10^7$ Monte Carlo steps
MCS. Then, the next $10^9$ MCS are used to compute averages. All
calculations were carried out using the parallel cluster BACO
located  at Instituto de F\'{\i}sica Aplicada, Universidad
Nacional de San Luis-CONICET, San Luis, Argentina. This facility
consists of 60 personal computers each with a 3.0 GHz Pentium-4
processor and 90 personal computers each with a 2.4 GHz Core 2
Quad processor.

In our Monte Carlo simulations, we set the temperature $T$, varied
the density $\theta=N/M$ and monitored the order parameter
$\delta$, which can be calculated as simple average. The
quantities related with the order parameter, such as the
susceptibility $\chi$, and the reduced fourth-order cumulant $U_L$
introduced by Binder~\cite{BINDER} were calculated as:
\begin{equation}
\chi = \frac{L^2}{k_BT} [ \langle \delta^2 \rangle - \langle
\delta \rangle^2] \label{chi}
\end{equation}
and
\begin{equation}
U_L = 1 -\frac{\langle \delta^4\rangle} {3\langle
\delta^2\rangle^2}, \label{ul}
\end{equation}
where $\langle \cdots \rangle$ means the average over the MC
simulation runs.

In addition, in order to discuss the nature of the phase
transition, the fourth-order energy cumulant $U_E$ was obtained as
\begin{equation}
U_E = 1 -\frac{\langle H^4\rangle} {3\langle H^2\rangle^2}.
\label{ue}
\end{equation}

\section{Results and conclusions}

As discussed in section I, the phase diagram of a system of
self-assembled rigid rods on square lattices has been recently
reported by Tavares et al. \cite{Tavares}. The authors showed that
the critical density, at which the IN transition occurs, increases
monotonically as $k_BT/w$ is increased. Thus, the nematic phase is
stable at low temperatures and high densities [see Fig. 1(a) in
Ref. [\onlinecite{Tavares}]]. Later, and based on this finding,
the nature of the IN transition at intermediate densities was
studied in Ref. [\onlinecite{PRE4}]. There, the calculations were
developed at $w=4 k_BT$ and different densities.

In the present study, we set the lateral interaction to $w=4.5
k_BT$ [a close value to that used in Ref. [\onlinecite{PRE4}]],
thus allowing a direct comparison with previous results for square
lattices. Accordingly, the density was varied around half
coverage. For each value of $\theta$, the effect of finite size
was investigated by examining lattices with $L$ ranging from $30$
to $180$.

\begin{figure}[t]
\includegraphics[width=6cm,clip=true]{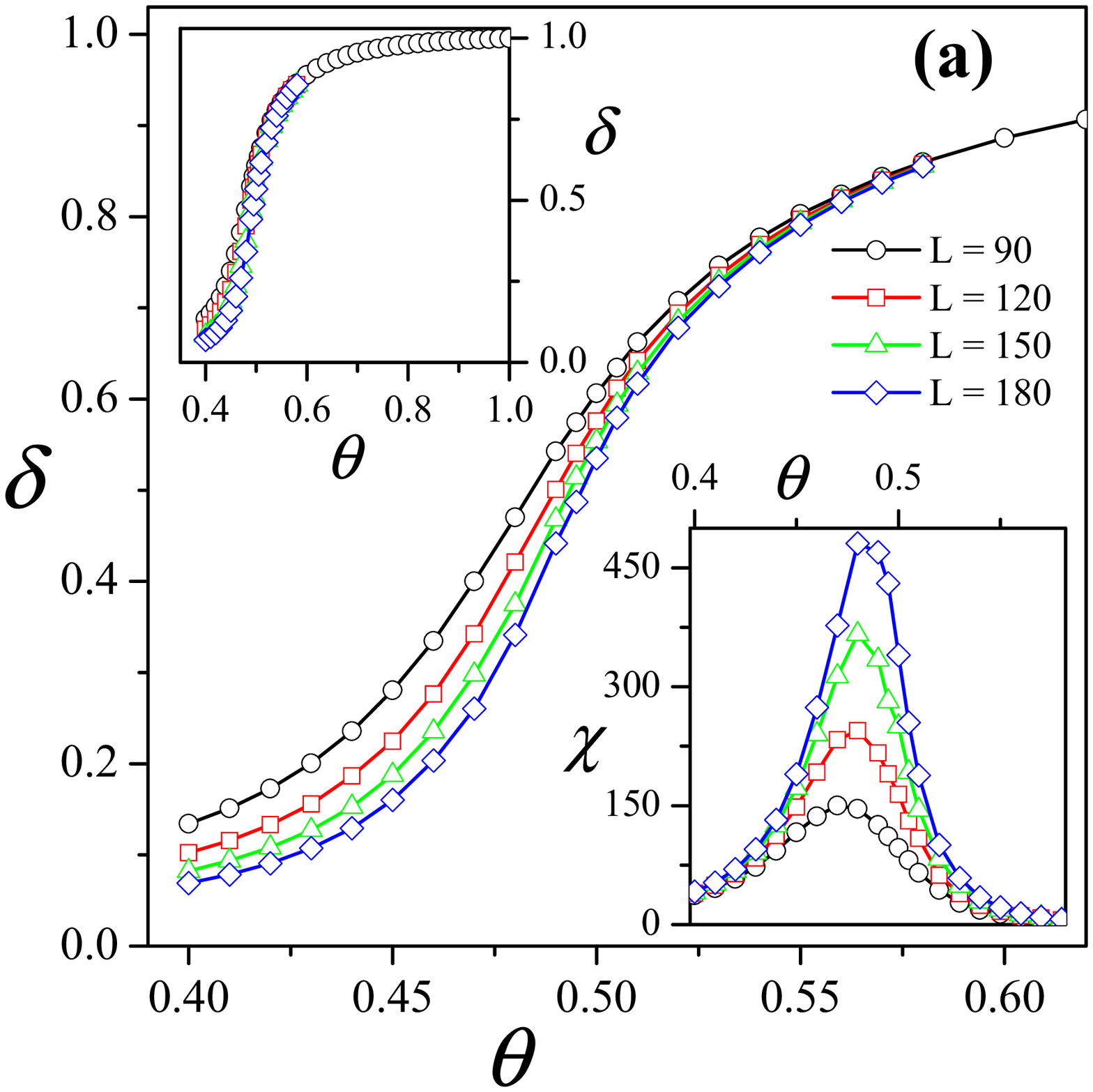}
\includegraphics[width=6cm,clip=true]{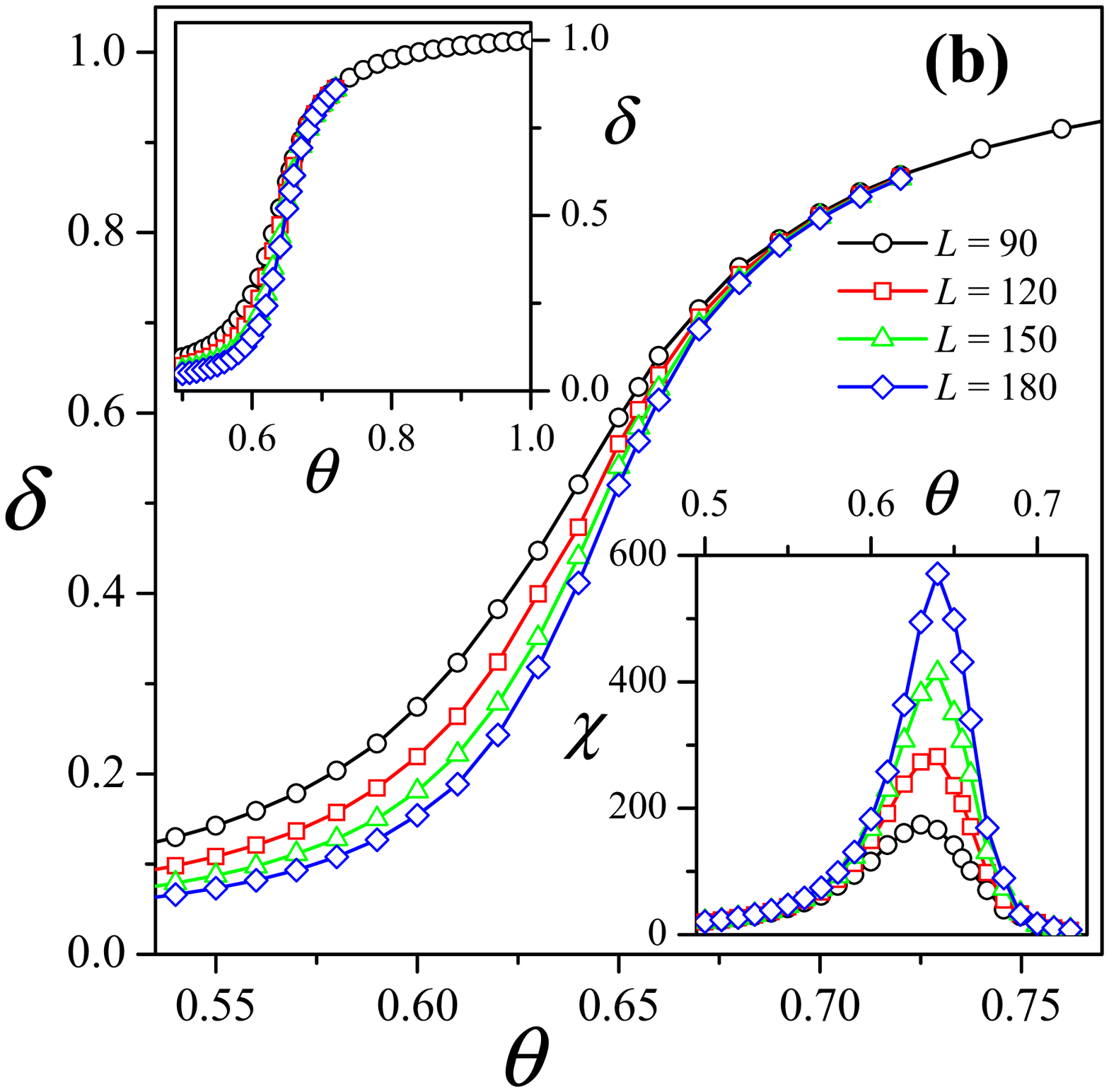}
\caption{Size dependence of the order parameter as a function of
coverage. Upper-left inset: Order parameter in the regime of high
densities. Lower-right inset: Size dependence of the
susceptibility as a function of coverage. The error in each
measurement is smaller than the size of the symbols. (a)
Triangular lattices. (b) Honeycomb lattices. } \label{figure2}
\end{figure}

We start with the calculation of the order parameter (Fig. 2), and
susceptibility (lower-right insets in Fig. 2) plotted versus
$\theta$ for several lattice sizes as indicated. Curves in part
(a) [(b)] correspond to triangular [honeycomb] lattices. As it can
be observed, $\delta$ appears as a proper order parameter to
elucidate the phase transition. When the system is disordered, all
orientations are equivalents and $\delta$ is zero. In the critical
regime, the particles align along one direction and $\delta$
increases continuously to one, remaining constant up to full
coverage (see upper-left insets in Fig. 2, where the order
parameter is shown up to $\theta=1$). In other words, nematic
order survives until $\theta=1$. This finding allows us to discard
the existence of a reentrant nematic transition at high densities
as speculated in Ref.~[\onlinecite{Tavares}]. With respect to the
susceptibility, the curves show a single peak which grows and
sharpens as the lattice size is increased.

\begin{figure}[t]
\includegraphics[width=6cm,clip=true]{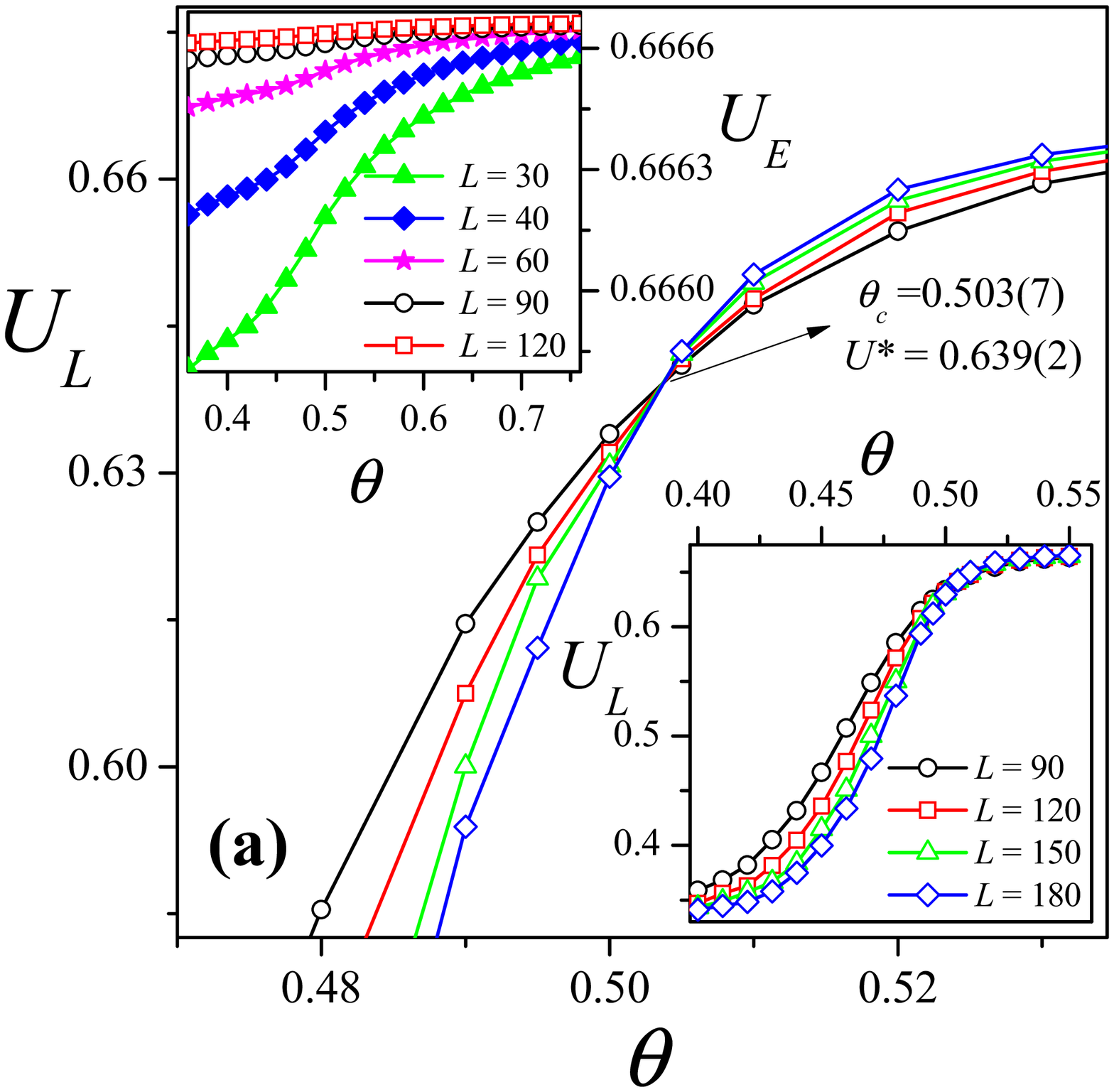}
\includegraphics[width=6cm,clip=true]{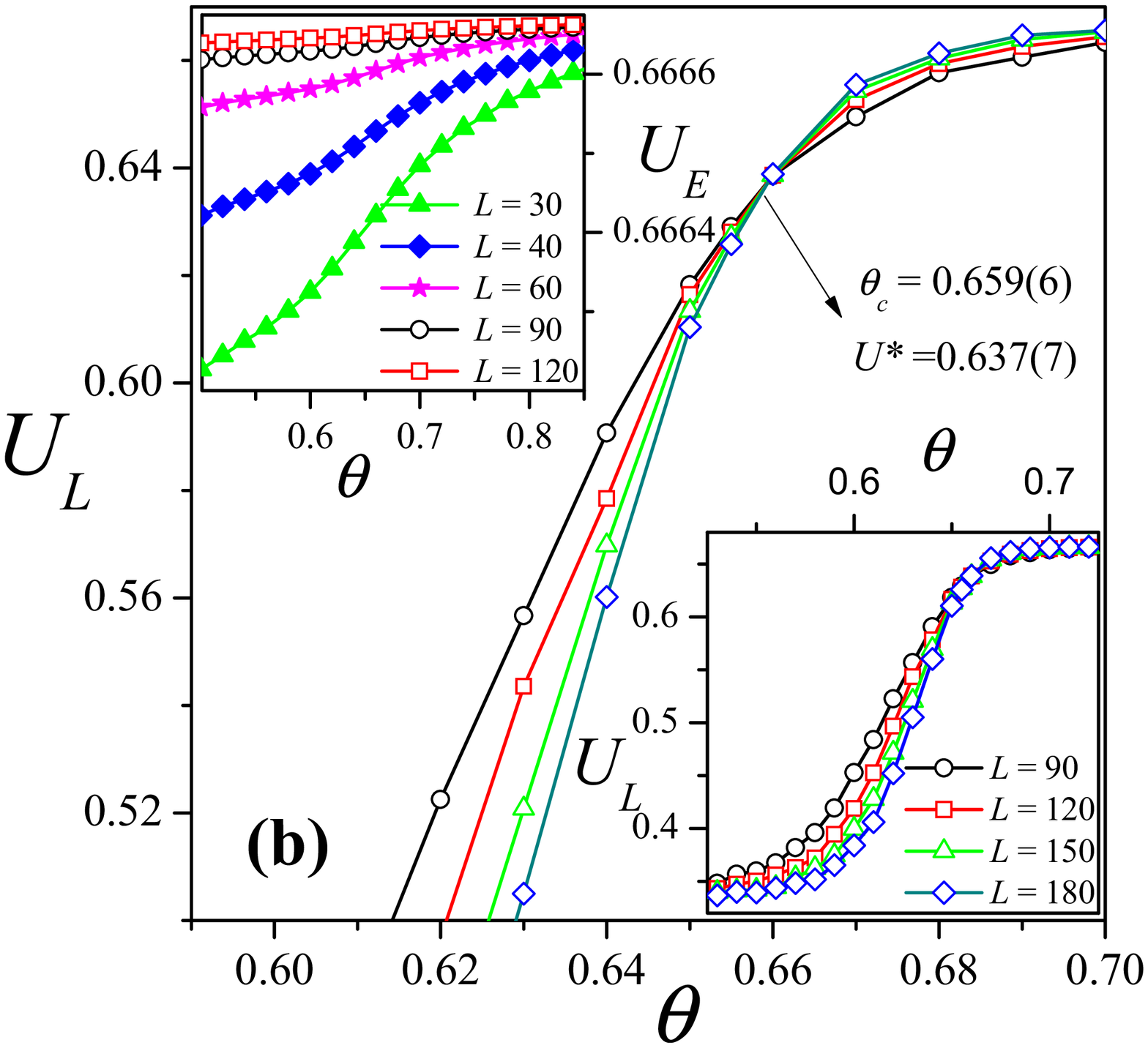}
\caption{Curves of $U_L(\theta)$ vs $\theta$ for lattices
of different sizes. From their intersections one obtained
$\theta_c$. In the lower-right inset, the data are plotted over a
wider range of densities. Upper-left inset: Coverage variation of
$U_E$ for various lattice sizes. (a) Triangular lattices. (b)
Honeycomb lattices. } \label{figure3}
\end{figure}

The critical density has been estimated from the plots of the
reduced fourth-order cumulants $U_L(\theta)$ plotted vs. $\theta$
for several lattice sizes (see Fig. 3). In the vicinity of the
critical point, cumulants show a strong dependence on the system
size. However, at the critical point the cumulants adopt a
nontrivial value $U^*$; irrespective of system sizes in the
scaling limit. Thus, plotting $U_L(\theta)$ for different linear
dimensions yields an intersection point $U^*$, which gives an
accurate estimation of the critical density in the infinite
system. In this case, the values obtained for the critical density
were $\theta_c=0.503(7)$ for triangular lattices [part (a)] and
$\theta_c=0.659(6)$ for honeycomb lattices [part (b)].

With respect to the value of the cumulant at the transition
density, which allows us to make a ``preliminary" identification
of the universality class of the transition
\cite{BINDER,foot,SELKE1,SELKE2}, we obtained $U^* =0.639(2)$ for
triangular lattices [part (a)] and $U^* =0.637(7)$ for honeycomb
lattices [part (b)]. These values of $U^*$ are practically
indistinguishable from previous estimates for the two-dimensional
one-state Potts model (see for instance Ref. [\onlinecite{PRE4}],
where $U^* =0.639(3)$). This result may be taken as a first
indication of universality. In the lower-right inset, the data are
plotted over a wider range of densities. As can be seen, the
curves exhibit the typical behavior of $U_L$ in the presence of a
continuous phase transition. Namely, the order-parameter cumulant
shows a smooth increase up to its ordered-phase value of $2/3$
instead of the characteristic deep (negative) minimum, as in a
first-order phase transition \cite{BINDER}.

In order to discard the possibility that the phase transition is a
first-order one, the energy cumulants [Eq. (9)]  have been
measured. As is well known, the finite-size analysis of $U_E$ is a
simple and direct way to determine the order of a phase transition
\cite{Binder2,Challa,Vilmayr}. Upper-left inset in Fig. 3
illustrates the energy cumulants plotted versus $\theta$ for
different lattice sizes. Data in part (a) correspond to triangular
lattices. Data in part (b) correspond to honeycomb lattices. As is
observed, $U_E$ has the characteristic behavior of a continuous
phase transition. Namely, the curves of $U_E$ show a dip close to
the critical density for all system sizes, but this minimum scales
to $2/3$ in the thermodynamic limit. This indicates that the
latent heat is zero in the thermodynamic limit, which reinforces
the arguments given in the paragraphs above.

The results obtained up to here allow us to confirm the existence
of a continuous phase transition at intermediate coverage and low
temperature. In addition, the evaluation of the fixed point value
of the cumulants indicates that, as in the case of square lattices
\cite{PRE4}, the observed phase transition belongs to the $q=1$
Potts universality class. To corroborate these findings, the
critical behavior of the present model has been investigated by
means of finite-size scaling (FSS) analysis. The FSS theory
implies the following behavior of $\delta$, $\chi$ and $U_L$ at
criticality:
\begin{equation}
\delta = L^{-\beta/\nu} \tilde \delta(L^{1/\nu} \epsilon),
\label{ds}
\end{equation}
\begin{equation}
\chi= L^{\gamma/\nu}\tilde \chi(L^{1/\nu} \epsilon) \label{chis}
\end{equation}
and
\begin{equation}
U_L=\tilde U_L(L^{1/\nu} \epsilon), \label{uls}
\end{equation}
for $L \rightarrow \infty$, $\epsilon \rightarrow 0$ such that
$L^{1/\nu} \epsilon $= finite, where $\epsilon \equiv
\theta/\theta_c - 1$. Here $\beta$, $\gamma$ and $\nu$ are the
standard critical exponents of the order parameter ($\delta \sim
-\epsilon^{\beta} $ for $\epsilon\rightarrow 0^-$, $L\rightarrow
\infty$),
 susceptibility($\chi \sim |\epsilon|^\gamma$ for $\epsilon \rightarrow 0$, $L\rightarrow \infty$) and
correlation length $\xi$ ($\xi \sim |\epsilon|^{-\nu}$ for
$\epsilon \rightarrow 0, L \rightarrow \infty$), respectively.
$\tilde \delta, \tilde \chi $ and $\tilde U_L$ are scaling
functions for the respective quantities.

\begin{figure}[t]
\includegraphics[width=6cm,clip=true]{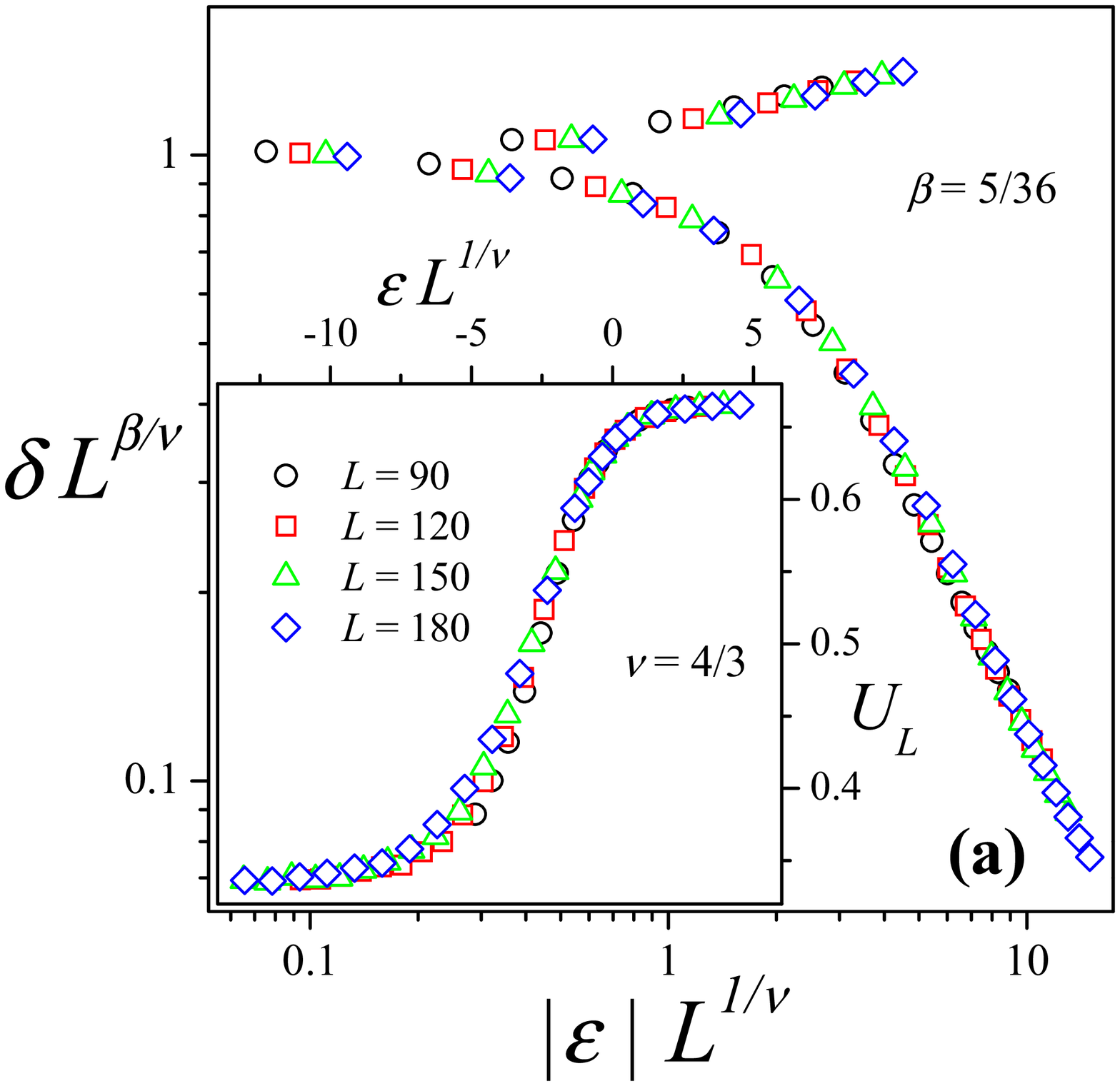}
\includegraphics[width=6cm,clip=true]{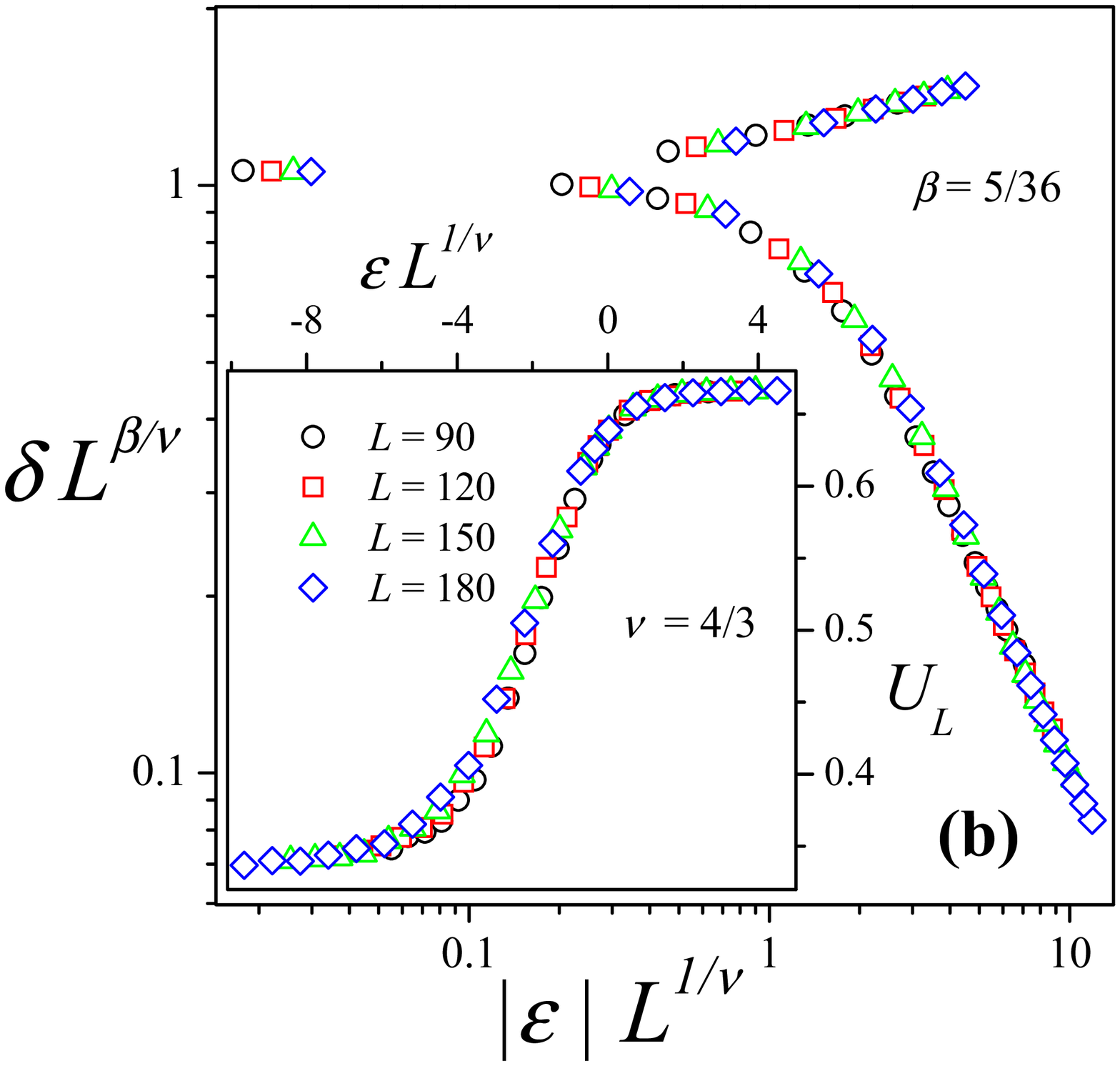}
\caption{(a) Data collapsing of the order parameter, $\delta
L^{\beta/\nu}$ vs $|\epsilon| L^{1/\nu}$, and of the cumulant,
$U_L$ vs $\epsilon L^{1/\nu}$ (inset) for triangular lattices. The
plots were made using $\theta_c=0.503$ and the exact percolation
exponents $\nu=4/3$ and $\beta=5/36$. (b) Same as part (a) for
honeycomb lattices and $\theta_c=0.659$. } \label{figure4}
\end{figure}

\begin{figure}[t]
\includegraphics[width=6cm,clip=true]{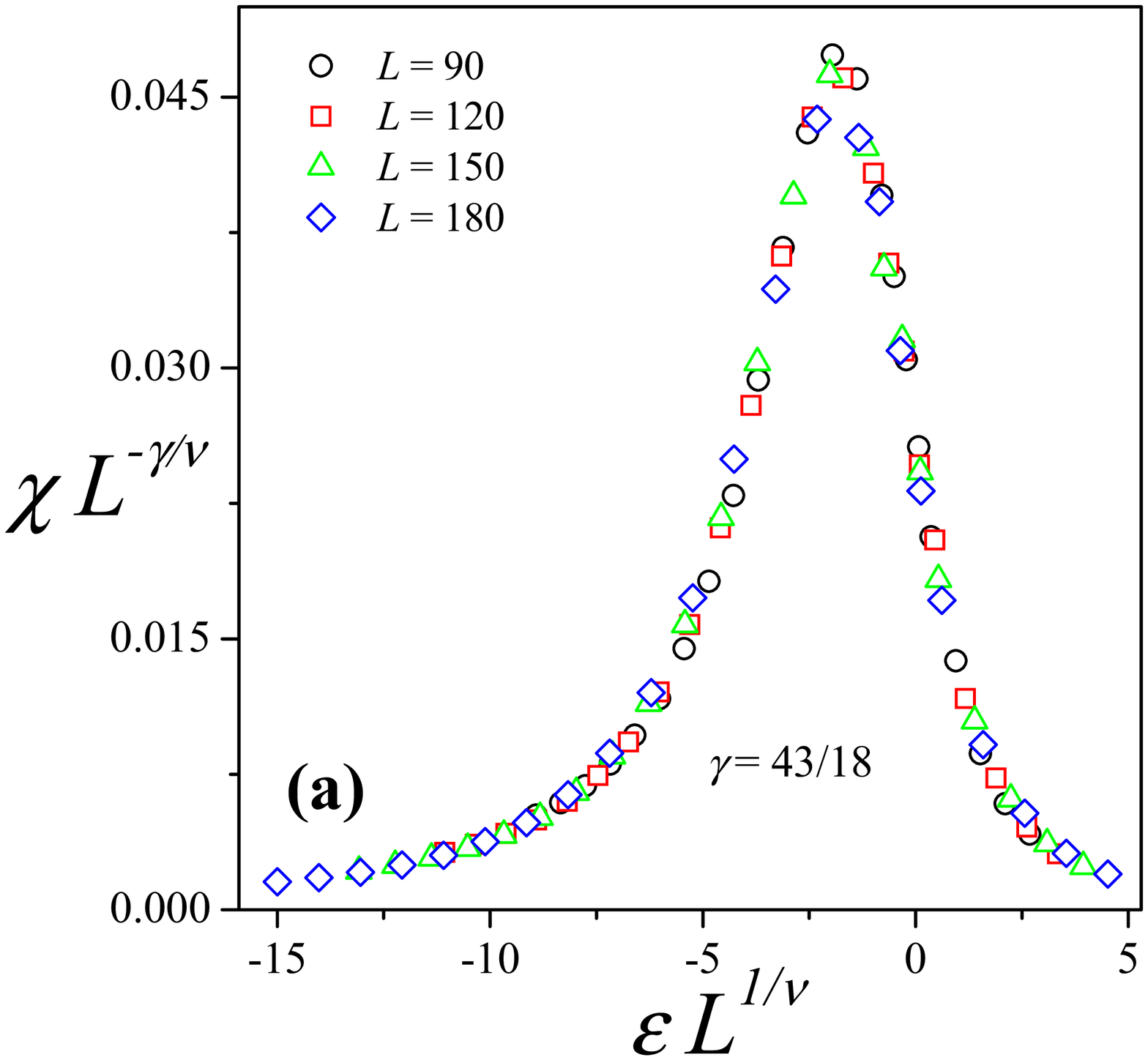}
\includegraphics[width=6cm,clip=true]{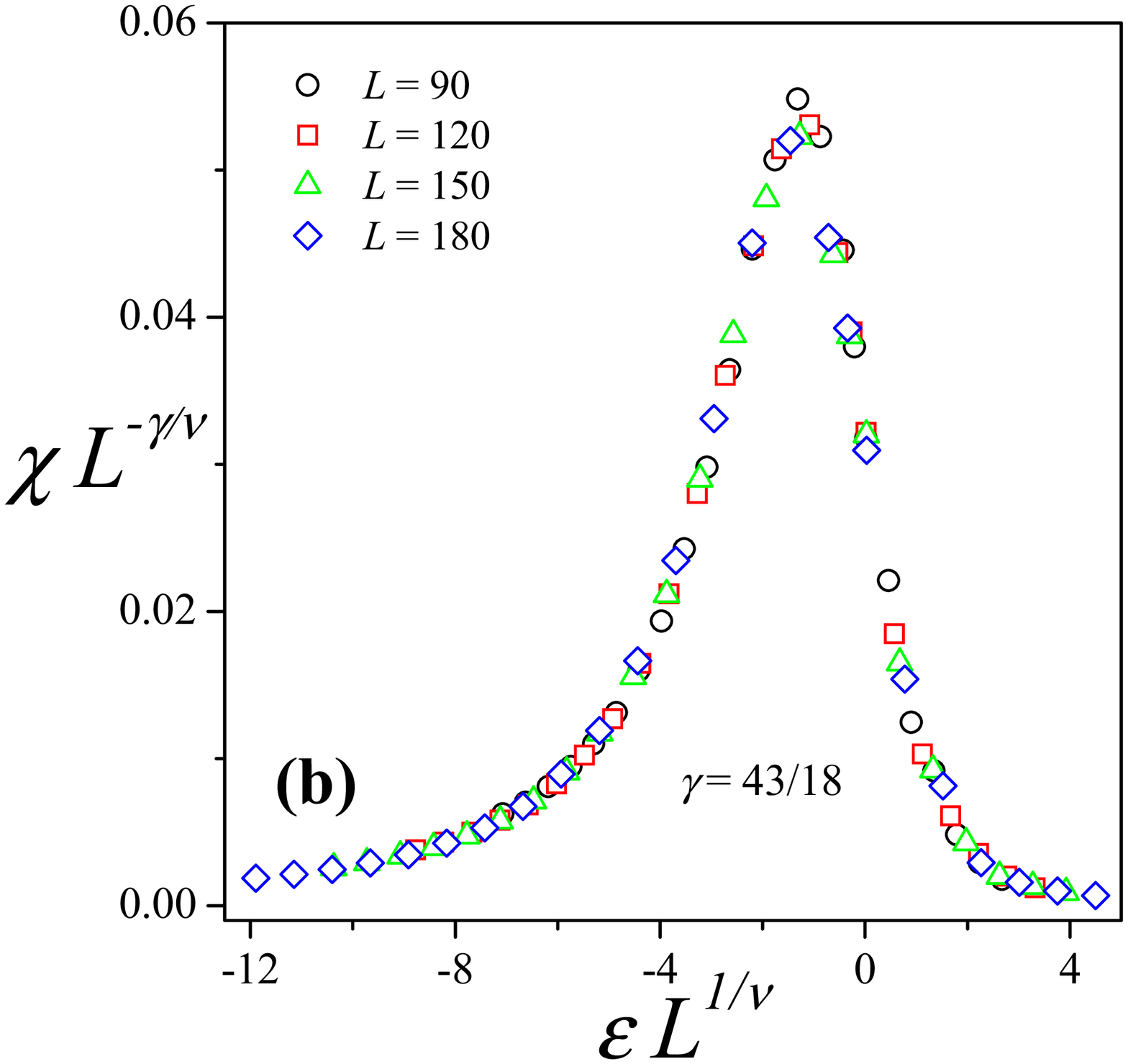}
\caption{ (a) Data collapsing of the the susceptibility, $\chi
L^{-\gamma/\nu}$ vs $\epsilon L^{1/\nu}$, for the curves in the
inset of Fig. 2 (a). The plot was made using $\theta_c=0.503$ and
the exact percolation exponents $\nu=4/3$ and $\gamma=43/18$. (b)
Same as part (a) for the curves in the inset of Fig. 2 (b) and
$\theta_c=0.659$. } \label{figure5}
\end{figure}

According to eqs.~(\ref{ds}-\ref{uls}), the scaling behavior can
be tested by plotting $\langle \delta \rangle L^{\beta/\nu}$ vs
$|\epsilon|L^{1/\nu}$, $\chi L^{-\gamma/\nu}$ vs $\epsilon
L^{1/\nu}$, and $U_L$ vs $\epsilon L^{1/\nu}$ and looking for data
collapsing. Using the exact values of the critical exponents of
the ordinary percolation (one-state Potts model) $\nu=4/3$,
$\beta=5/36$, and $\gamma=43/18$, we obtain an excellent scaling
collapse as it is shown in Figs. 4 and 5. Data in part (a)
correspond to triangular lattices and $\theta_c=0.503$. Data in
part (b) correspond to honeycomb lattices and $\theta_c=0.659$.
The study in Figs. 4 and 5 corroborates that the transition
observed corresponds to the universality class of the $q=1$ Potts
model.

Finally, it is important to remark that the universality observed
for self-assembled rigid rods on triangular and honeycomb lattices
coincides with that reported in Ref. [\onlinecite{PRE4}] for
square lattices. In other words, the universality class of the
present model does not depend on the lattice geometry. This result
$(1)$ reinforces the the arguments given in Ref.
[\onlinecite{PRE4}] linking the thermal phase transition (IN phase
transition) occurring in the system and the percolation behavior
of the clusters of aligned monomers \cite{foot1}; and $(2)$
reveals a significant difference between the model here discussed
and that of monodisperse rigid rods adsorbed on two-dimensional
lattices with a discrete number of orientations
\cite{GHOSH,EPL1,PHYSA19}, where the universality class of the IN
phase transition occurring at intermediate densities belongs to
the 2D Ising universality class for square lattices and to the
three-state Potts universality class for triangular and honeycomb
lattices.

In summary, we have used Monte Carlo simulations and finite-size
scaling theory to study the critical properties of self-assembled
rigid rods on triangular and honeycomb lattices at intermediate
density. The existence of a IN continuous phase transition was
confirmed. In addition, the scaling behavior of the system
revealed that, as in the case of square lattices, the phase
transition belongs to the $q=1$ Potts universality class.

\acknowledgments This work was supported in part by CONICET
(Argentina) under project number PIP 112-200801-01332; Universidad
Nacional de San Luis (Argentina) under project 322000 and the
National Agency of Scientific and Technological Promotion
(Argentina) under project 33328 PICT 2005.




\begin{thebibliography}{99}
\bibitem{foot1} The appearance of nematic order at critical density is accompanied
by the simultaneous presence of a percolating cluster of aligned
monomers connecting the extremes of the lattice.
\end{thebibliography}
\end{document}